\documentclass[aps,prl,preprint,preprintnumbers,superscriptaddress,amsmath,amssymb]{revtex4}

\usepackage{subfigure}

\usepackage{color}
\usepackage{graphicx}
\usepackage{dcolumn}
\usepackage{bm}

\usepackage{natbib}

\begin{document}

\preprint{DESY 11-057}

\title{Anomalous scaling in the random-force-driven Burgers equation:\\A Monte Carlo study}

\author{D. Mesterh\'azy}
 \email{mesterh@crunch.ikp.physik.tu-darmstadt.de}
 \affiliation{Institut f\"ur Kernphysik, Technische Universit\"at Darmstadt\\Schlossgartenstra\ss e 9/2,
64289 Darmstadt, Germany}
\author{K. Jansen}
 \email{Karl.Jansen@desy.de}
 \affiliation{Deutsches Elektronen-Synchrotron (DESY)\\Platanenallee 6,
15738 Zeuthen, Germany}

\date{\today}

\begin{abstract}
We present a new approach to determine numerically the statistical behavior of small-scale structures in hydrodynamic turbulence. Starting from the functional integral representation of the random-force-driven Burgers equation we show that Monte Carlo simulations allow us to determine the anomalous scaling of high-order moments of velocity differences. Given the general applicability of Monte Carlo methods, this opens up the possibility to address also other systems relevant to turbulence within this framework.
\end{abstract}

\maketitle

The small-scale statistical properties of hydrodynamic turbulence is an old and tantalizing problem \cite{1995tlan.book.....F}. For turbulent flow stirred at large scales and far from the boundaries one expects a universal scaling for the small-scale fluctuations. Indeed, experiment gives strong indications for such universal behavior in Navier-Stokes turbulence \cite{1997AnRFM..29..435S,2002PhFl...14.1065G,2008PhRvL.100w4503B,2009AnRFM..41..165I,2010JFM...653..221B}. The exact values of the scaling exponents however are still under debate. In such a situation it is useful to have a model system at hand that shares some essential properties with the original problem and allows for a clear physical understanding.

The random-force-driven Burgers equation
\begin{equation}
\partial_{t} u + u \partial_{x} u - \nu \partial_{x}^{2} u = f(x,t)~,
\label{burgers}
\end{equation}
is one such example. It was originally conceived as a one-dimensional model for compressible hydrodynamic turbulence \cite{burgers1973nonlinear} and provides a useful benchmark setting to test new analytical and numerical methods for real-world turbulence \cite{2000nlin.....12033F,2007PhR...447....1B}. Here, $u$ is the velocity, and $f$ a centered random field displaying Gaussian statistics. We will consider the special case where the system is driven by a self-similar forcing that is white in time. The two-point correlation function of the stochastic forcing in Fourier space is given by
\begin{equation}
\langle f(k ,t ) f(k' ,t') \rangle \propto D_{0} |k|^{\beta} \delta(k+k') \delta(t-t')~,
\label{forcing}
\end{equation}
where the parameter $\beta$ determines the relative importance of the stirring mechanism at different scales, and the dimensionful constant $D_{0}$ measures its strength. For $\beta$ large and negative the forcing effectively acts at large scales. On the other hand, the kinematic viscosity $\nu$ in \eqref{burgers} provides a dissipation scale $\eta$ and for $\nu \rightarrow 0^{+}$ the two characteristic scales $\eta$, and the system size $L$ separate. The stochastic forcing drives the system into a non-equilibrium steady state, where in the range \mbox{$\eta \ll |k|^{-1} \ll L$} the energy flux through wavenumber $k$ behaves as \mbox{$\Pi(k) \propto |k|^{1 + \beta}$} \cite{1996PhRvE..54.4681H,1997PhRvE..56.4259H}. 

\begin{figure}[!t]
\centering
\includegraphics[width=0.45\textwidth]{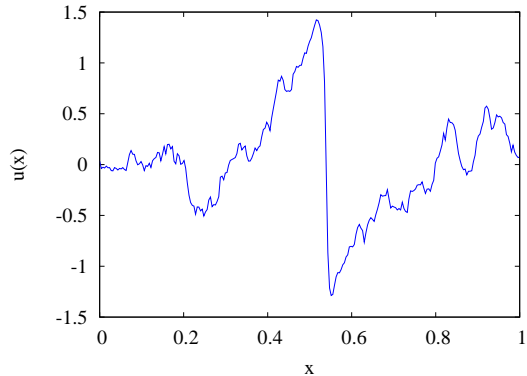}
\caption{\label{fig:Profile} Typical velocity profile $u(x)$ from a simulation on a \mbox{$254 \times 1024$} (space $\times$ time) lattice, where $x$ is taken in units of the spatial lattice size $L$.}
\end{figure}

The case $\beta = -1$ corresponds to the physically interesting situation where the flux $\Pi(k)$ is constant (up to logarithmic corrections) and the interplay of the stochastic forcing and advective term leads to a Kolmogorov energy spectrum \mbox{$E(k) \propto |k|^{-5/3}$} \cite{1995PhRvE..51.2739C,1995PhRvE..52.5681C}. The physical picture behind this scenario is the appearance of shocks with a finite dissipative width (see e.g. Fig.~\ref{fig:Profile}). The large fluctuations associated with the negative gradient of the front give the dominant contribution to the anomalous scaling of velocity differences $\Delta u = u(x+r) - u(x)$. In particular, we have the structure functions $\langle | \Delta u |^{n} \rangle \propto r^{\zeta_{n}}$, and the scaling exponents $\zeta_{n} = 1$ for $n \geq 3$ strongly deviate from the Kolmogorov scaling prediction $\zeta_{n} = n/3$ that follows from a naive dimensional analysis \cite{2000nlin.....12033F,2007PhR...447....1B}. These rare fluctuations are strongly non-Gaussian and lead to the known asymptotic left tail of the probability distribution function (PDF) for velocity differences $\mathcal{P}(\Delta u,r)$ \cite{1997PhRvL..78.1452B}.

Here, we approach the problem from the functional integral point of view \cite{1976ZPhyB..23..377J,1977JPhA...10..777P,1981JSP....25..183J}. The functional integral gives a non-perturbative definition of the field theory and thus, it is ideally suited to study the strong and rare fluctuations present in fully developed turbulence that give the main contribution to the high-order moments of velocity differences. By sampling the associated probability distribution functional via Monte Carlo methods we show that it is possible to determine the scaling behavior of structure functions from first principles. Monte Carlo simulations are directly transferable to other systems of interest and are free of any modeling assumptions. Though not directly competitive with conventional time-advancing methods as, e.g. pseudo-spectral or finite-difference methods, Monte Carlo simulations may provide a unique perspective on such important problems as, e.g. intermittency in fully developed turbulence \cite{1997PhRvL..78.1452B}. In view of the well-established anomalous scaling behavior of Burgers turbulence \cite{2005PhRvL..94s4501M,2007PhR...447....1B} and the physical picture of the underlying mechanisms for intermittency \cite{1995PhRvE..52.6183P,1996PhRvE..54.4908G,1997PhRvL..78.1452B}, this provides an ideal setting to test our method and understand possible systematic effects at finite Reynolds number and system size. We emphasize again that in this paper we are not aiming to complete, or even improve on the accuracy obtained with other methods for Burgers turbulence. We rather want to provide a test of the generally applicable functional integral method and to demonstrate that a very reasonable accuracy can be obtained from this approach, a fact that was highly unclear at the beginning of this project.

The functional integral for the random-force-driven Burgers equation is obtained via the Martin-Siggia-Rose formalism \cite{1973PhRvA...8..423M,1976ZPhyB..23..377J,1977JPhA...10..777P,1978PhRvB..18..353D,1981JSP....25..183J} by introducing an auxiliary response field $\mu$. We have the field theory 
\begin{equation}
Z = \int [d u]\, [d \mu] \exp \{ -S [u , \mu]\}~,
\end{equation}
with the action
\begin{eqnarray}
S &=& -i \int\! dt\, dx\, \mu ( \partial_t u + u \partial_x u - \nu \partial_x^2 u ) \nonumber \\
&& + \, \frac{1}{2} \int\! dt\, dx\, dy \, \mu(x,t) D(x - y) \mu(y,t) ~,
\end{eqnarray}
%\begin{equation}
%S = -i \int\! dt\, dx\, \mu ( \partial_t u + u \partial_x u - \nu \partial_x^2 u ) + \frac{1}{2} \int\! dt\, dx\, dy \, \mu(x,t) D(x - y) \mu(y,t) ~.
%\end{equation}
where $D(x-y)$ is the spatial part of the two-point correlation function \eqref{forcing}. In this form the action does not satisfy positivity. To obtain a Gibbs measure that can be sampled by a Markov chain Monte Carlo (MCMC) algorithm we integrate out the auxiliary field. This leaves us with the probability density functional
\begin{eqnarray}
P[ u ] &=& \exp \Big\{ - \frac{1}{2} \int\! dt\, dx\, dy \, ( \partial_t u + u \partial_x u - \nu \partial_x^2 u )  \nonumber \\ 
&& D^{-1}(x-y) ( \partial_t u + u \partial_x u - \nu \partial_x^2 u ) \Big\}~.
\label{action}
\end{eqnarray}
%\begin{equation}
%P[ u ] = \exp \Big\{ - \frac{1}{2} \int\! dt\, dx\, dy \, ( \partial_t u + u \partial_x u - \nu \partial_x^2 u ) D^{-1}(x-y) ( \partial_t u + u \partial_x u - \nu \partial_x^2 u ) \Big\}
%\label{action}
%\end{equation}
The theory is then defined by placing the field $u(x,t)$ on the sites of a regular space-time lattice $\Lambda$, i.e. $(x,t) \in \Lambda$. This way, we impose a UV cutoff that eliminates the details of those processes occurring deep in the dissipative regime. Then, the measure is given by \mbox{$[d u] \,\rightarrow \prod_{(x,t) \,\in\, \Lambda}\! d u(x,t)$} and the action in \eqref{action} needs to be discretized appropriately. We replace the dynamics \eqref{burgers} with a finite-difference equation with backward-time discretization
\begin{equation}
\partial_t u + u \partial_x u \rightarrow \frac{1}{\epsilon} (u(t) - u(t-\epsilon)) + u(t-\epsilon) \,\partial_{x} u(t-\epsilon)~,
\end{equation}
where $\epsilon$ is the lattice spacing in time direction. This ensures the correct dynamics in the continuum limit \cite{ZinnJustin:2002ru}. For the advective term we take the anti-symmetric spatial derivative
\begin{equation}
\partial_{x} u \rightarrow \frac{1}{2 a} ( u(x+a) - u(x-a))~,
\end{equation}
where $a$ is the lattice spacing in the spatial direction. With this choice of discretization the problem is amenable to a local over-relaxation algorithm \cite{2008EL.....8440002D}. Starting from an initial configuration $\{ u(x,t) , (x,t) \in \Lambda \}$ the set of single-site variables is updated iteratively by the successive application of a transition probability $P(u(x,t) \rightarrow u'(x,t))$. We use the high-quality \verb|ranlux| (pseudo) random number generator \cite{1994CoPhC..79..100L} which is essential for large-scale lattice simulations. Specific improvements, e.g. Chebyshev acceleration \cite{1962mia..book.....V} significantly reduce thermalization and autocorrelation times for the relevant observables in the inertial range.

It is necessary to map the discretized theory to its continuum counterpart and one has to ensure that the parameters are well-defined in the continuum limit. For that purpose the kinematic viscosity is identified with $\nu = \hat{\nu} \, a^2 / \epsilon$ where $\hat{\nu}$ is the viscosity in lattice units, and the Reynolds number scales as $\textrm{Re} \propto \nu^{-1}$. Furthermore, we have to ensure that the relevant scales of the system are resolved. In particular, we have to ensure that the dissipation scale fits on the lattice, i.e. $\eta = \textrm{Re}^{-3/4} L \gtrsim a$ where $L$ is the IR scale present in our system as a consequence of the finite lattice size. One may immediately recognize that this imposes a hard constraint on the realization of lattice simulations -- fully developed turbulence requires a large computational effort where the number of lattice sites in the spatial direction scales as $\propto \textrm{Re}^{3/4}$, for given $L$. In practice, we are therefore bound to work at non-zero viscosity $\nu$.

\begin{figure*}[!t]
\subfigure{\includegraphics[width=0.45\textwidth]{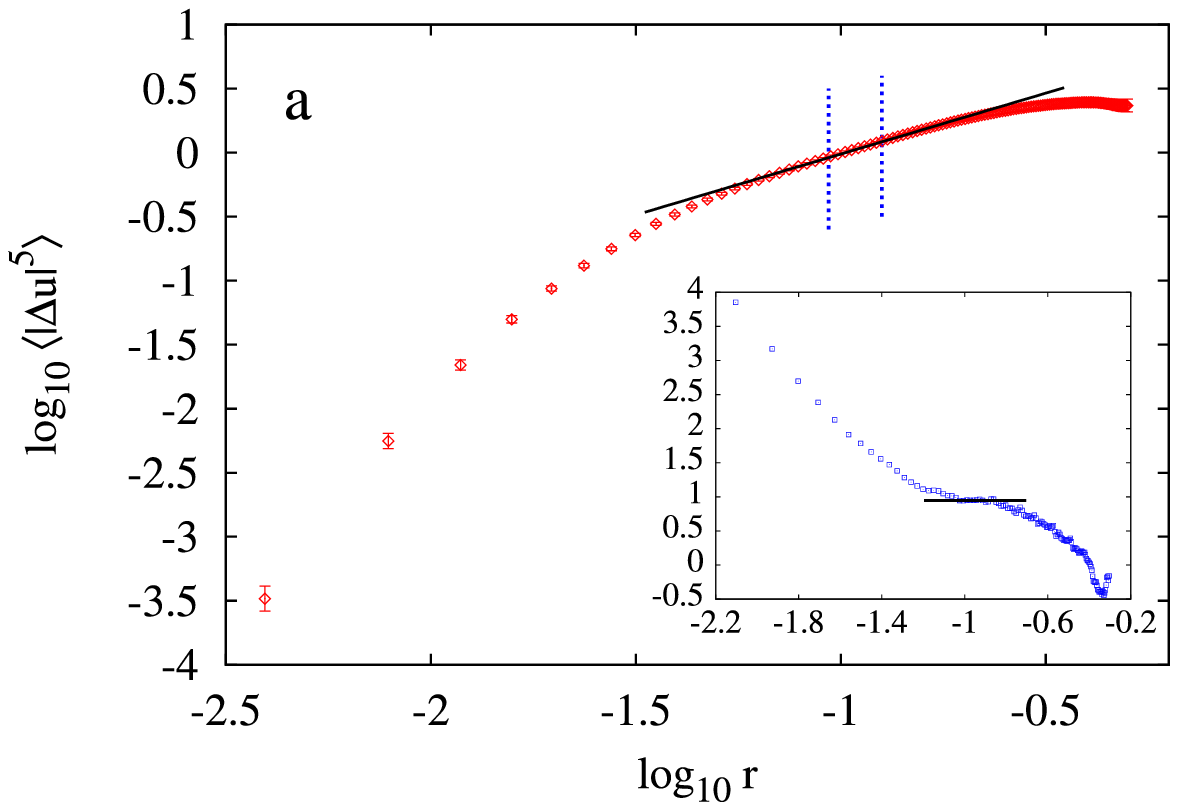}}
\subfigure{\includegraphics[width=0.45\textwidth]{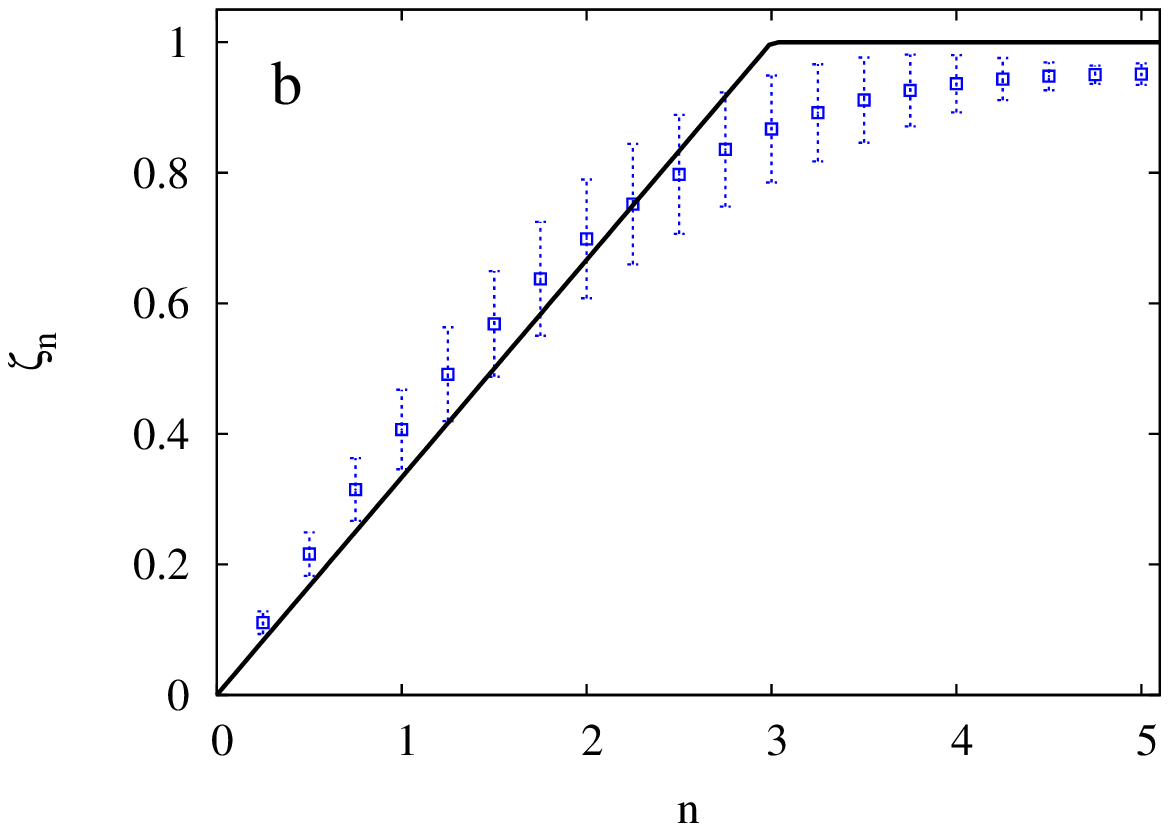}}
\caption{\label{fig:StructureFunction} (a) Log-log plot of the structure function of order $n = 5$ with a linear scaling function plotted for comparison. Vertical bars indicate the region for the extraction of scaling exponents. Inset shows the local slopes versus $r$. (b) Structure function scaling exponents $\zeta_n$ versus order $n$. The black curve indicates a bifractal scaling behavior.}
\end{figure*}

Simulations at moderate to high Reynolds numbers require massively parallel architectures. State of the art simulations at $\textrm{Re} = 64$ and lattice size $254 \times 1024$ (space $\times$ time direction) run on up to $512$ processors in parallel. Structure functions are evaluated over an ensemble of configurations generated by the MCMC algorithm as described in the previous paragraphs. For every configuration we measure velocity differences from a randomly chosen starting point. This dramatically reduces autocorrelations for our sample. In Fig.~\ref{fig:StructureFunction}a we show, as an example, the $n=5$ order structure function calculated for an ensemble from a $254 \times 1024$ lattice simulation, averaged over nearly $5 \times 10^{5}$ statistically independent field configurations. To determine the scaling range \emph{a priori} is difficult, and a well-known problem in the literature (see e.g. \cite{1995tlan.book.....F}). Here, we employ a working definition where it is defined as the range of scales that minimizes the $\chi^2$ of a linear least-squares (LLS) fit to the fifth order structure function in the log-log plot. The corresponding region is indicated in Fig.~\ref{fig:StructureFunction}a. For comparison we have included the values of the local slope (evaluated over three consecutive points) in the inset. We identify a plateau where the local exponents are nearly constant -- this defines the value of the scaling exponent. In general, with this method, we cannot rule out subleading terms or possible logarithmic corrections that may influence the scaling behavior \cite{2004NJPh....6...37B,2004PhRvL..92i4503B,2005PhRvL..94s4501M}. In fact, such a situation is very likely and can lead to the appearance of multiscaling \cite{2005PhRvL..94s4501M}. While in principle these contributions should be taken into account for the accurate determination of the scaling behavior, in practice it is difficult to distinguish different types of scaling contributions without any further assumptions. We obtain the scaling spectrum (Fig.~\ref{fig:StructureFunction}b) where the error bars given are those of the LLS fit in the scaling range. Clearly, the $n = 5$ data point in Fig.~\ref{fig:StructureFunction}b has minimal error which follows simply from our definition of the scaling range. We see that the scaling exponents are close to the bifractal scaling prediction \cite{2000nlin.....12033F,2007PhR...447....1B}, and within error bars agrees with the results of \cite{2005PhRvL..94s4501M}, obtained at high spectral resolution. As a last remark we want to add that we have not used extended self-similarity (ESS) \cite{1996PhRvE..53.3025B} at any point in our analysis. Though ESS may enlarge the effective scaling range we found that it can suggest a clean scaling behavior even if subleading terms are present.

Since we are dealing with a finite system both in space and time one may expect boundary effects. In our simulations we have chosen periodic boundary conditions in space and fixed (Dirichlet) boundary conditions in time. This way we eliminate zero mode effects from the dynamics. For a space-time lattice of infinite extent in the time direction the probability measure \eqref{action} defines a stationary process, i.e. correlation functions will only depend on time differences. We have checked this property explicitly in our analysis -- far from the boundaries, in the bulk of the configurations, the system is approximately in a stationary state.

In the continuum both the action in \eqref{action} and the measure are invariant under the set of
Galilean transformations
\begin{equation}
x \rightarrow x + r ~, \quad u(x,t) \rightarrow u(x + r,t) + v ~, \quad r = v t ~.
\label{GT}
\end{equation}
To avoid an over counting of physically equivalent field configurations one should eliminate these modes by the Faddeev-Popov procedure \cite{ZinnJustin:2002ru}. While gauge fixing is unavoidable for generic correlators \cite{2007PhRvL..99y4501B,2009NuPhB.814..522B} this is not so for velocity differences, as solely considered in this work which are manifestly invariant under transformations \eqref{GT}.

\begin{figure*}[!t]
\centering
\subfigure{\includegraphics[width=0.45\textwidth]{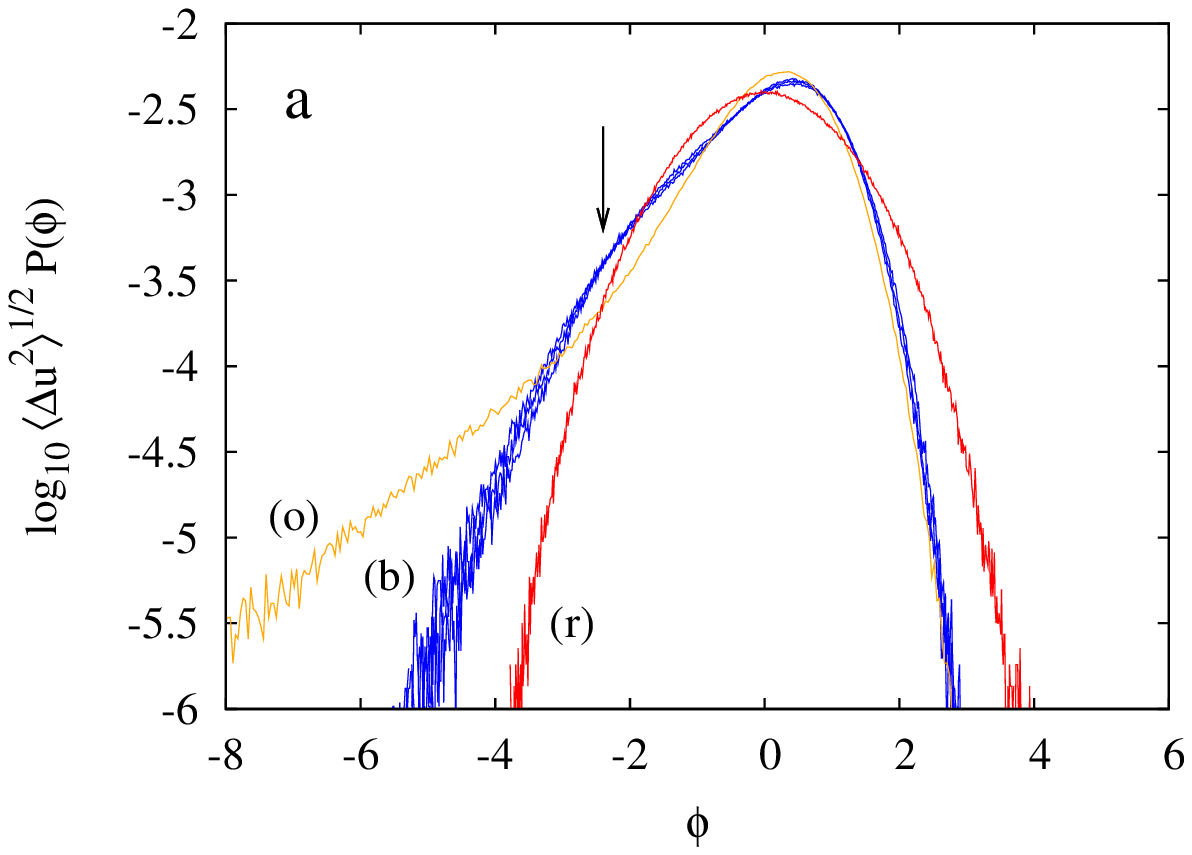}}\qquad
\subfigure{\includegraphics[width=0.45\textwidth]{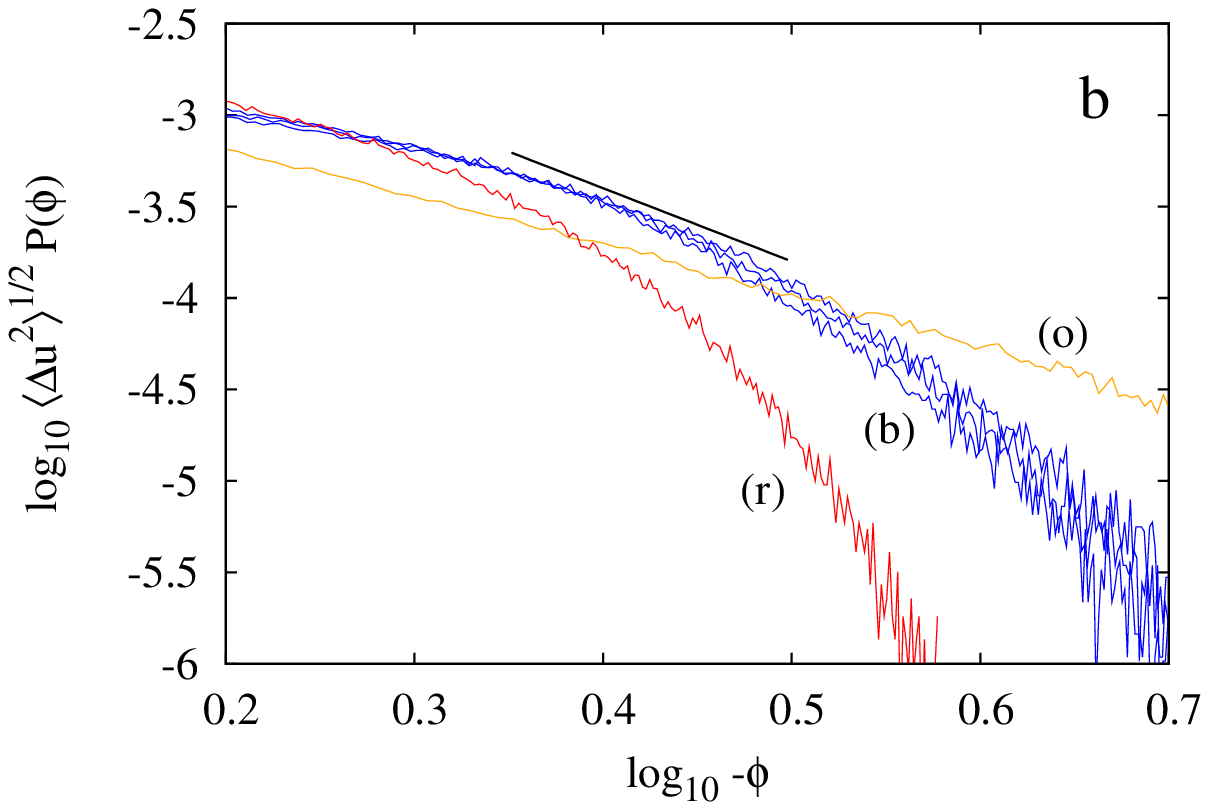}}
\caption{\label{fig:PDF1} Probability distribution functions $\mathcal{P}(\Delta u , r)$ as a function of the dimensionless variable $\phi = \Delta u / [ \langle \Delta u^{2} \rangle ]^{1/2}$ plotted for different values of $r$. (a) Collapse of the PDF in the universal regime (blue). In the energy-containing range (red) the fluctuations become Gaussian -- the random forcing dominates -- whereas in the dissipative regime (orange) fluctuations are strongly enhanced. (b) Scaling region for the left tail of the PDF. The black line indicates the scaling prediction with exponent $\gamma = -4$.}
\end{figure*}

One may also check the statistics for velocity differences directly on the level of the probability distribution functions $\mathcal{P}(\Delta u , r)$. This gives valuable qualitative information on the physical behavior in our simulations of Burgers turbulence. In Fig.~\ref{fig:PDF1}a we show the PDF of velocity differences for a set of values of the separation $r$, where we use the dimensionless variable $\phi = \Delta u / [ \langle \Delta u^{2} \rangle ]^{1/2}$ to quantify the fluctuations. At large scales, far from the inertial range we clearly recognize the effects of the random Gaussian forcing (red). In the dissipative region the left tail of the PDF is especially pronounced and captures the strong fluctuations described by the shocks (orange). For separations $\eta \ll r \ll L$ in the inertial range we see that the PDF $\mathcal{P}(\Delta u , r )$, plotted for three different values of $r$, nicely collapse onto each other (blue). In particular, in the regime where the fluctuations are much smaller than the root-mean-square velocity $|\Delta u| \ll u_{rms}$, the PDF of velocity differences has a universal scaling form
\begin{equation}
\mathcal{P} (\Delta u , r) = r^{-z} f(\Delta u / r^{z})~,
\end{equation}
where $z$ is the dynamic exponent \cite{1996PhRvL..77.3118Y}. In the asymptotic region $-\Delta u / r^{z} \gg 1 $ where $\Delta u < 0$ we expect the algebraic scaling $\mathcal{P}(\Delta u,r) \propto (\Delta u)^{{\gamma}}$ with exponent $\gamma = -4$. The relevant region is shown in Fig.~\ref{fig:PDF1}a (indicated by the arrow) and Fig.~\ref{fig:PDF1}b. The corresponding scaling prediction with exponent $\gamma = -4$ is plotted for comparison as the black line in Fig.~\ref{fig:PDF1}b. Though our statistics are not sufficient to give a tight prediction on the scaling exponent, indications for the conjectured scaling behavior can be inferred from Fig.~\ref{fig:PDF1}b.

At this point we want to give a short remark on some issues that arise when turning to incompressible three-dimensional Navier-Stokes turbulence. It is well-known, that the inclusion of the pressure term is one of the main obstacles in simulations of turbulence, as the requirement of incompressibility introduces non-local interactions. In the functional integral formulation this leads to a non-vanishing Fadeev-Popov determinant that can be treated by standard procedures (see e.g. \cite{Rothe:2005nw}). In our lattice approach, we have chosen to use a local update over-relaxation algorithm that proved to be quite efficient for our purposes. The long-range correlations in our system imposed by the forcing however, prohibit any attempt to parallelize in the spatial direction. This poses a severe problem when turning to higher dimensions. For that purpose it is absolutely mandatory to switch to a global update algorithm as, e.g. a Hybrid Monte Carlo algorithm \cite{1987PhLB..195..216D} that is usually used in standard lattice calculations when non-local actions are considered. The implementation of non-trivial spatial boundary conditions is an interesting possibility when turning to higher dimensions. In principle, there are no restrictions on the type of boundary conditions in our simulations. However, in such a case the two-point correlation function of the forcing will not be restricted by symmetry arguments and may take a rather difficult form. 

The Burgers equation provides an ideal setting to understand systematic effects at finite Reynolds numbers and finite lattice sizes in the framework of Monte Carlo simulations. We have demonstrated that our simulations are able to reproduce the well-known anomalous scaling behavior in the Burgers model. It is important to remark that this is possible without exploiting the integrability property of the Burgers equation, as was done e.g. with a fast Legendre transform algorithm in \cite{1994JScCom...9..259P,2005PhRvL..94s4501M}. Thus, Monte Carlo simulations are directly applicable to other physical systems of interest where it is important to have alternative methods available to determine the statistical behavior of small-scale fluctuations. Certainly, the numerical efficiency of our method is an issue, and there is room for improvement. Specifically, one may ask if a global Hybrid Monte Carlo algorithm will improve the performance of our simulations. Another question isolated from the numerical efficiency relates to the evaluation of the scaling spectrum and the role of subleading scaling corrections. For that matter it is essential to distinguish possibly different types of scaling contributions. Further work in this direction is in progress. This work presents a first important step towards the determination of scaling behavior for the case of full three-dimensional Navier-Stokes turbulence in the framework of Monte Carlo simulations.

\begin{acknowledgments}

We want to thank Gernot M\"unster and Dirk Homeier for useful discussions. Furthermore, D.M. wants to thank Claudio Rebbi for the invitation and an inspiring stay at Boston University. We have performed simulations on the IBM p690 cluster \emph{Jump} and the BlueGene/P system \emph{JuRoPa} at Forschungszentrum J\"ulich, and a significant part of the work was done on the BlueGene/L system at the Center of Computational Science at Boston University.

\end{acknowledgments}

\bibliography{references}

\end{document}